\documentclass[final,5p]{elsarticle}

\usepackage{amssymb}
\usepackage{algorithm}
\usepackage{algpseudocode}
\usepackage{xcolor}
\usepackage{amsmath}
\usepackage{booktabs}
\usepackage{adjustbox}
\usepackage{hyperref}

\journal{ArXiv (a short version was accepted by ICSE'24 poster track)} 

\begin{document}

\begin{frontmatter}



\title{Graph Neural Networks based \\ Log Anomaly Detection and Explanation}


\author[inst1]{Zhong Li\corref{cor}\fnref{label1}}
\fntext[label1]{Corresponding author.}
\ead{z.li@liacs.leidenuniv.nl}
\author[inst1]{Jiayang Shi}
\ead{j.shi@liacs.leidenuniv.nl}
\author[inst1]{Matthijs van Leeuwen}
\ead{
m.van.leeuwen@liacs.leidenuniv.nl}

\affiliation[inst1]{organization={LIACS, Leiden University},%Department and Organization
            %addressline={Address Two}, 
            city={Leiden},
            %postcode={22222}, 
            %state={State Two},
            country={the Netherlands}}

\begin{abstract}
  Event logs are widely used to record the status of high-tech systems, making log anomaly detection important for monitoring those systems. Most existing log anomaly detection methods take a log event count matrix or log event sequences as input, exploiting quantitative and/or sequential relationships between log events to detect anomalies. However, only considering quantitative or sequential relationships may result in low detection accuracy. To alleviate this problem, we propose a graph-based method for unsupervised log anomaly detection, dubbed \textit{Logs2Graphs}, which first converts event logs into attributed, directed, and weighted graphs, and then leverages graph neural networks to perform graph-level anomaly detection. Specifically, we introduce One-Class Digraph Inception Convolutional Networks, abbreviated as OCDiGCN, a novel graph neural network model for detecting graph-level anomalies in a collection of attributed, directed, and weighted graphs. By integrating graph representation and anomaly detection, OCDiGCN learns a specialized representation that leads to high detection accuracy. Crucially, we furnish a concise set of nodes pivotal in OCDiGCN's prediction as explanations for each detected anomaly, offering valuable insights for subsequent root cause analysis. Experiments on five benchmark datasets show that \textit{Logs2Graphs} exhibits comparable or superior performance when compared to state-of-the-art log anomaly detection methods.
  %By coupling the graph representation and anomaly detection steps, OCDiGCN can learn a representation that is especially suited for anomaly detection, resulting in a high detection accuracy. 
  % Importantly, for each identified anomaly, we additionally provide a small subset of nodes that play a crucial role in OCDiGCN's prediction as explanations, which can offer valuable cues for subsequent root cause diagnosis.
  %Experiments on five benchmark datasets show that \textit{Logs2Graphs} performs at least on par with state-of-the-art log anomaly detection methods on simple datasets while largely outperforming state-of-the-art log anomaly detection methods on complicated datasets.
\end{abstract}


\end{frontmatter}


\section{Introduction}
Modern high-tech systems, such as cloud computers or lithography machines, typically consist of a large number of components. Over time these systems have become larger and more complex, making manual system operation and maintenance hard or even infeasible \cite{li2022feature}. Therefore, automated system operation and maintenance is highly desirable. To achieve this, system logs are universally used to record system states and important events. By analysing these logs, faults and potential risks can be identified, and remedial actions may be taken to prevent severe problems. System logs are usually semi-structured texts though, and identifying anomalies through log anomaly detection is often challenging.  

Since both industry and academia show  great interest in identifying anomalies from logs, a plethora of log anomaly detection methods have been proposed. Existing log anomaly detection methods can be roughly divided into three categories: quantitative-based, sequence-based, and graph-based methods. Specifically, quantitative-based methods, such as OCSVM \cite{scholkopf2001estimating} and PCA \cite{xu2009detecting}, utilise a log event count matrix to detect anomalies, and are therefore unable to capture semantic information of and sequential information between log events. Meanwhile, sequence-based methods, including DeepLog \cite{du2017deeplog} and LogAnomaly \cite{meng2019loganomaly}, aim to detect anomalies by taking sequential (and sometimes semantic) information into account. They cannot consider the full structure among log events though. In contrast, graph-based methods, such as GLAD-PAW \cite{wan2021glad} and GLAD \cite{li2023glad}, convert logs to graphs and exploit semantic information as well as the structure among log events, exhibiting the following three advantages over the former two categories of methods \cite{jia2017approach}: 1) they are able to identify problems for which the structure among events is crucial, such as performance degradation; 2) they are capable of providing contextual log messages corresponding to the identified problems; and 3) they can provide the `normal' operation process in the form of a graph, helping end-users find root-causes and take remedial actions. However, graph-based methods like GLAD transform log events into \textit{undirected} graphs though, which may fail to capture important information on the order among log events. Moreover, most existing graph-based methods perform graph representation and anomaly detection separately, leading to suboptimal detection accuracy. 

{\color{black}Moreover, as highlighted by Li et al. \cite{li2022survey}, with the growing adoption of anomaly detection algorithms in safety-critical domains, there is a rising demand for providing explanations for the decisions made within those domains. This requirement, driven by ethical considerations and regulatory mandates, underscores the importance of accountability and transparency in such contexts. Moreover, in practical applications, the attainment of precise anomaly explanations contributes to the timely isolation and diagnosis of anomalies, which can mitigate the impact of anomalies by facilitating early intervention \cite{yang2021semi}. However, to our knowledge, most existing log anomaly detection methods focus exclusively on accurate detection without giving  explanations.}

To overcome these limitations, we propose \textit{Logs2Graphs}, a graph-based unsupervised log anomaly detection approach for which we design a novel one-class graph neural network. Specifically, \textit{Logs2Graphs} first utilises off-the-shelf methods to learn a semantic embedding for each log event, and then assigns log messages to different groups. Second, \textit{Logs2Graphs} converts each group of log messages into an attributed, directed, and weighted graph, with each node representing a log event, the node attributes containing its semantic embedding, directed edges representing how an event is followed by other events, and the corresponding edge weights indicating the number of times the events follow each other. Third, by coupling the graph representation learning and anomaly detection objectives, we introduce One-Class Digraph Inception Convolutional Networks (OCDiGCN) as a novel method to detect anomalous graphs from a set of graphs. As a result, \textit{Logs2Graphs} leverages the rich and expressive power of attributed, directed, and edge-weighted graphs to represent logs, followed by using graph neural networks to effectively detect graph-level anomalies, taking into account both semantic information of log events and structure information (including sequential information as a special case) among log events. {\color{black}Importantly, by decomposing the anomaly score of a graph into individual nodes and visualizing these nodes based on their contributions, we provide understandable explanations for identified anomalies.}

Overall, our contributions can be summarised as follows: (1) We introduce \textit{Logs2Graphs}, which formalises log anomaly detection as a graph-level anomaly detection problem and represents log sequences as directed graphs to capture more structure information than previous approaches; (2) We introduce OCDiGCN, a general end-to-end unsupervised graph-level anomaly detection method for attributed, directed and edge-weighted graphs. By coupling the graph representation and anomaly detection objectives, we improve the potential for accurate anomaly detection over existing approaches; (3) {\color{black}For each detected anomaly, we identify important nodes as explanations, offering cues for subsequent root cause diagnosis;} (4) We empirically compare our approach to eight state-of-the-art log anomaly detection methods on five benchmark datasets, showing that \textit{Logs2Graphs} performs at least on par with and often better than its competitors.

%The reminder of this paper is organised as follows. Section 2 revisits related work, after which Section 3 formalises the problem. Section 4 describes Digraph Inception Convolutional Networks~\cite{tong2020digraph}, which are used for \textit{Logs2Graphs} in Section 5. We then evaluate \textit{Logs2Graphs} in Section 6 and conclude in Section 7.

\section{Related Work}
Graph-based log anomaly detection methods usually comprise five steps: log parsing, log grouping, graph construction, graph representation learning, and anomaly detection. In this paper we focus on graph representation learning, log anomaly detection, and explanation, thus only revisiting related work in these fields. 
%Specifically, graph representation learning and log anomaly detection can be performed in a two-stage manner (namely graph-level representation learning followed by anomaly detection) or in an end-to-end manner (namely graph-level representation for anomaly detection). 

\subsection{Graph Representation Learning}

Graph-level representation learning methods, such as GIN \cite{xu2018powerful} and Graph2Vec \cite{narayanan2017graph2vec}, are able to learn a mapping from graphs to vectors. Further, graph kernel methods, including Weisfeiler-Lehman (WL) \cite{shervashidze2011weisfeiler} and Propagation Kernels (PK) \cite{neumann2016propagation}, can directly provide pairwise distances between graphs. Both types of methods can be combined with off-the-shelf anomaly detectors, such as OCSVM \cite{scholkopf2001estimating} and iForest \cite{liu2008isolation}, to perform graph-level anomaly detection.

To improve on these na\"{i}ve approaches, efforts have been made to develop graph representation learning methods especially for anomaly detection. For instance,  OCGIN \cite{zhao2021using} and GLAM \cite{zhao2022graph} combine the GIN \cite{xu2018powerful} representation learning objective with the SVDD objective \cite{tax2004support} to perform graph-level representation learning and anomaly detection in an end-to-end manner. GLocalKD \cite{ma2022deep} performs random distillation of graph and node representations to learn `normal' graph patterns. Further, OCGTL \cite{qiu2022raising} combines neural transformation learning and one-class classification to learn graph representations for anomaly detection. Although these methods are unsupervised or semi-supervised, they can only deal with attributed, undirected, and unweighted graphs. 

iGAD \cite{zhang2022dual} considers graph-level anomaly detection as a graph classification problem and combines attribute-aware graph convolution and substructure-aware deep random walks to learn graph representations. However, iGAD is a supervised method, and can only handle attributed, undirected, and unweighted graphs. CODEtect \cite{nguyen2020anomaly} takes a pattern-based modelling approach using the minimum description length principle and identifies anomalous graphs based on \textit{motifs}. CODEtect can (only) deal with labelled, directed, and edge-weighted graphs, but is computationally very expensive. In contrast, we introduce a general unsupervised method for graph-level anomaly detection that can handle attributed, directed and edge-weighted graphs.

\subsection{Log Anomaly Detection and Explanation}
Log anomaly detection methods can be roughly divided into: 1) traditional, `shallow' methods, such as principal component analysis (PCA) \cite{xu2009detecting}, one-class SVM (OCSVM) \cite{scholkopf2001estimating}, isolation forest (iForest) \cite{liu2008isolation}, and histogram-based outlier score (HBOS) \cite{goldstein2012histogram}, which take a log event count matrix as input and analyse quantitative relationships; 2) deep learning based methods, such as DeepLog \cite{du2017deeplog}, LogAnomaly \cite{meng2019loganomaly}, and AutoEncoder \cite{farzad2020unsupervised}, which employ sequences of log events (and sometimes their semantic embeddings) as input, analysing sequential information and possibly semantic information of log events to identify anomalies; and 3) graph-based methods, such as TCFG \cite{jia2017approach} and GLAD-PAW \cite{wan2021glad}, which first convert logs into graphs and then perform graph-level anomaly detection. 

To our knowledge, only a few works~\cite{wan2021glad,zhang2022deeptralog,xie2022loggd, li2023glad} have capitalised on the powerful learning capabilities of graph neural networks for log anomaly detection. GLAD \cite{li2023glad} transforms logs into undirected, weighted, and attributed heterogeneous graph, and propose a temporal-attentive graph edge anomaly detection model to detect anomalous relations. However, converting logs into undirected graphs may result in loss of important sequential information. Further, DeepTraLog \cite{zhang2022deeptralog} combines traces and logs to generate a so-called Trace Event Graph, which is an attributed and directed graph. On this basis, they train a Gated Graph Neural Networks based Deep Support Vector Data Description model to identify anomalies. However, their approach requires the availability of both traces and logs, and is unable to handle edge weights. In contrast, like LogGD  \cite{xie2022loggd} and GLAD-PAW \cite{wan2021glad}, our proposed \textit{Logs2Graphs} approach is applicable to generic logs by converting logs into attributed, directed, and edge-weighted graphs. However, LogGD is a supervised method that requires fully labelled training data, which is usually impractical and even impossible. Moreover, LogGD and GLAD-PAW \cite{wan2021glad} are not capable of providing explanations for identified anomalies. In contrast, our proposed algorithm OCDiGCN is a general \textit{unsupervised} graph-level anomaly detection method for attributed, directed, and edge-weighted graphs, able to provide explanations for identified anomalies.

{\color{black}Although anomaly explanation has received much attention in traditional anomaly detection \cite{li2022survey,li2023QCAD}, only a few studies \cite{yang2021plelog} considered log anomaly explanation. Specifically, PLELog \cite{yang2021plelog} offers explanations by quantifying the significance of individual log events within an anomalous log sequence, thereby facilitating improved identification of relevant log events by operators. Similarly, our method provides explanations for anomalous log groups by identifying and visualising a small subset of important nodes.}

\section{Problem Statement}
Before we state the log anomaly detection problem, we first introduce notation and definitions regarding event logs and graphs.

\smallskip \noindent 
\textbf{Event logs}. \textit{Logs} are used to record system status and important events, and are usually collected and stored centrally as log files. A \textit{log file} typically consists of many \textit{log messages}. Each \textit{log message} is composed of three components: a timestamp, an event type (\textit{log event} or \textit{log template}), and additional information (\textit{log parameters}). \textit{Log parsers} are used to extract log events from log messages.

Further, log messages can be grouped into \textit{log groups} (a.k.a. \textit{log sequences}) using certain criteria. Specifically, if a \textit{log identifier} is available for each log message, one can group log messages based on such identifiers. Otherwise, one can use a \textit{fixed} or \textit{sliding window} to group log messages. %The \textit{window size} can be determined according to timestamp or the number of observations. %For example, one can take the log messages within a hour as a group or take every 100 log messages as a group.
Besides, counting the occurrences of each log event within a log group yields an \textit{event count vector}. Consequently, for a log file consisting of many log groups, one can obtain an \textit{event count matrix}. The process of generating an \textit{event count matrix} (or other feature matrix) is known as \textit{feature extraction}. Extracted features are often used as input to an anomaly detection algorithm to identify \textit{log anomalies}, i.e., log messages or log groups that deviate from what is considered `normal'.

\smallskip \noindent 
\textbf{Graphs}. We consider an attributed, directed, and edge-weighted graph  $\mathcal{G} = (\mathcal{V},  \mathcal{E},  \textbf{X}, \textbf{Y})$, where $\mathcal{V} = \{v_{1},...,v_{|\mathcal{V}|}\}$ denotes the set of \textit{nodes} and $\mathcal{E} = \{e_{1},...,e_{|\mathcal{E}|}\} \subseteq \mathcal{V}\times\mathcal{V}$ represents the set of edges. If $(v_{i},v_{j})\in \mathcal{E}$, then there is an edge from node $v_{i}$ to node $v_{j}$. Moreover, $\textbf{X} \in \mathbb{R}^{|\mathcal{V}|\times d}$ is the node attribute matrix, with the $i$-th row representing the attributes of node $v_{i}$, and $d$ is the number of attributes. Besides, $\textbf{Y} \in \mathbb{N}^{|\mathcal{E}|\times |\mathcal{E}|}$ is the edge-weight matrix, where $\textbf{Y}_{ij}$ represents the weight of the edge from node $v_{i}$ to $v_{j}$.

Equivalently, $\mathcal{G}$ can be described as $(\textbf{A}, \textbf{X}, \textbf{Y})$, with adjacency matrix $\textbf{A} \in \mathbb{R}^{|\mathcal{V}|\times|\mathcal{V}|}$, where $\textbf{A}_{ij} = \mathbb{I}[(v_{i},v_{j}) \in \mathcal{E}]$ indicates whether there is an edge from node $v_{i}$ to node $v_{j}$, for $i,j \in \{1,...,|\mathcal{V}|\}$. 

\subsection{Graph-based Log Anomaly Detection}

Given a set of log files, we let $\mathcal{L} = \{L_{1},...,L_{|\mathcal{L}|}\}$ denote the set of unique log events. We divide the log messages into $M$ log groups $\textbf{Q} = \{ \textbf{q}_{1},...,\textbf{q}_{m},...,\textbf{q}_{M}\}$, where $\textbf{q}_{m}=\{\textbf{q}_{m1},...,\textbf{q}_{mn},...,\textbf{q}_{mN}\}$ is a log group and $\textbf{q}_{mn}$ a log message.

For each log group $\textbf{q}_{m}$, we construct an attributed, directed, and edge-weighted graph $\mathcal{G}_{m} = (\mathcal{V}_{m},  \mathcal{E}_{m},  \textbf{X}_{m}, \textbf{Y}_{m})$ to represent the log messages and their relationships. Specifically, each node $v_{i} \in \mathcal{V}_{m}$ corresponds to exactly one log event $L \in \mathcal{L}$ (and vice versa). Further, an edge $e_{ij} \in \mathcal{E}_{m}$ indicates that log event $i$ is at least once immediately followed by log event $j$ in $\textbf{q}_{m}$. Attributes $\textbf{x}_{i} \in \textbf{X}_{m}$ represent the semantic embedding of log event $i$, and $y_{ij} \in \textbf{Y}_{m}$ is the weight of edge $e_{ij}$, representing the number of times event $i$ was immediately followed by event $j$. In this manner, we construct a set of log graphs $\{\mathcal{G}_{1},...,\mathcal{G}_{m},...,\mathcal{G}_{M}\}$.

We use these definitions to define graph-based log anomaly detection:

\smallskip \noindent 
\textbf{Problem 1 (Graph-based Log Anomaly Detection)}. \textit{Given a set of attributed, directed, and weighted graphs that represent logs, find those graphs that are notably different from the majority of graphs.}

\smallskip
What we mean by `notably different' will have to be made more specific when we define our method, but we can already discuss what types of anomalies can potentially be detected. Most methods aim to detect two types of anomalies:
\begin{itemize}
	\item A log group (here a graph) is considered a \textit{quantitative anomaly} if the occurrence frequencies of some events in the group are higher or lower than expected from what is commonly observed. For example, if a file is opened (event $A$) twice, it should normally also be closed (event $B$) twice. In other words, the number of event occurrences $\#A = \#B$ in a normal pattern and an anomaly is detected if $\#A \neq \#B$.
	\item A log group is considered to contain \textit{sequential anomalies} if the order of certain events violates the normal order pattern. For instance, a file can be closed only after it has been opened in a normal workflow. In other words, the order of event occurrences $A \rightarrow B$ is considered normal while $B \rightarrow A$ is considered anomalous.
\end{itemize} 

An advantage of graph-based anomaly detection is that it can detect these two types of anomalies, but also anomalies reflected in the structures of the graphs. Moreover, no \textit{unsupervised} log anomaly detection approaches represent event logs as attributed, directed, weighted graphs, which allow for even higher expressiveness than undirected graphs (and thus limiting the information loss resulting from the representation of the log files as graphs).

\section{Preliminaries: Digraph Inception Convolutional Nets}

To learn node representations for attributed, directed, and edge-weighted graphs, Tong et al. \cite{tong2020digraph} proposed Digraph Inception Convolutional Networks (DiGCN).%, which can obtain wider receptive fields and explore multi-scale features through diagraph convolution and $k$-th order proximity.

Specifically, given a graph $\mathcal{G}$ described by an adjacency matrix $\textbf{A} \in \mathbb{R}^{|\mathcal{V}|\times|\mathcal{V}|}$, a node attribute matrix $\textbf{X} \in \mathbb{R}^{|\mathcal{V}|\times d}$, and an edge-weight matrix $\textbf{Y} \in \mathbb{R}^{|\mathcal{V}|\times|\mathcal{V}|}$, DiGCN defines the $k$-th order digraph convolution as
\begin{equation}
	\textbf{Z}^{(k)} =
	\begin{cases}
		\textbf{X}\Theta^{(0)} & \text{$k=0$}\\
		\Psi\textbf{X}\Theta^{(1)}& \text{$k=1$}\\
		\Phi\textbf{X}\Theta^{(k)} & \text{$k \geq 2$},
	\end{cases}       
\end{equation}
where $\Psi=\frac{1}{2}\left(\Pi^{(1)\frac{1}{2}}\textbf{P}^{(1)}\Pi^{(1)\frac{-1}{2}}+\Pi^{(1)\frac{-1}{2}}\textbf{P}^{(1)T}\Pi^{(1)\frac{1}{2}}\right)$ and $\Phi = \textbf{W}^{(k)\frac{-1}{2}}\textbf{P}^{(k)}\textbf{W}^{(k)\frac{-1}{2}}$. Particularly, $\textbf{Z}^{(k)} \in \mathcal{R}^{|\mathcal{V}|\times f }$ denotes the convolved output with $f$ output dimension, and  $\Theta^{(0)},\Theta^{(1)},\Theta^{(k)}$ represent the trainable parameter matrices.

Moreover, $\textbf{P}^{(k)}$ is the $k$-th order proximity matrix defined as
\begin{equation}
	\textbf{P}^{(k)} =
	\begin{cases}
		\textbf{I} & \text{$k=0$}\\
		\tilde{\textbf{D}}^{-1}\tilde{\textbf{A}}& \text{$k=1$}\\
		Ins\left((\textbf{P}^{(1)})^{(k-1)}(\textbf{P}^{(1)T})^{(k-1)}\right)& \text{$k \geq 2$},
	\end{cases}       
\end{equation}
where $\textbf{I} \in \mathcal{R}^{|\mathcal{V}|\times |\mathcal{V}|}$ is an identity matrix,  $\tilde{\textbf{A}} = \textbf{A} + \textbf{I}$, and $\tilde{\textbf{D}}$ denotes the diagonal degree matrix with $\tilde{\textbf{D}}_{ii} = \sum_{j}\tilde{\textbf{A}}_{ij}$. Besides, $Ins\left((\textbf{P}^{(1)})^{(k-1)}(\textbf{P}^{(1)T})^{(k-1)}\right)$ is defined as  $$\frac{1}{2}Intersect\left((\textbf{P}^{(1)})^{(k-1)}(\textbf{P}^{(1)T})^{(k-1)},(\textbf{P}^{(1)T})^{(k-1)}(\textbf{P}^{(1)})^{(k-1)}\right)$$, with $Intersect(\cdot)$ denoting the element-wise intersection of two matrices (see \cite{tong2020digraph} for computation details). In addition, $\textbf{W}^{(k)}$ is the diagonalized weight matrix of  $\textbf{P}^{(k)}$, and $\Pi^{(1)}$ is the approximate diagonalized eigenvector of $\textbf{P}^{(1)}$. Particularly, the approximate diagonalized eigenvector is calculated based on personalised PageRank \cite{bahmani2010fast}, with a parameter $\alpha$ to control the degree of conversion from a digraph to an undirected graph. We omit the details to conserve space, and refer to \cite{tong2020digraph} for more details.

\begin{figure*}[h]
	\centering
	\includegraphics[width=13cm]{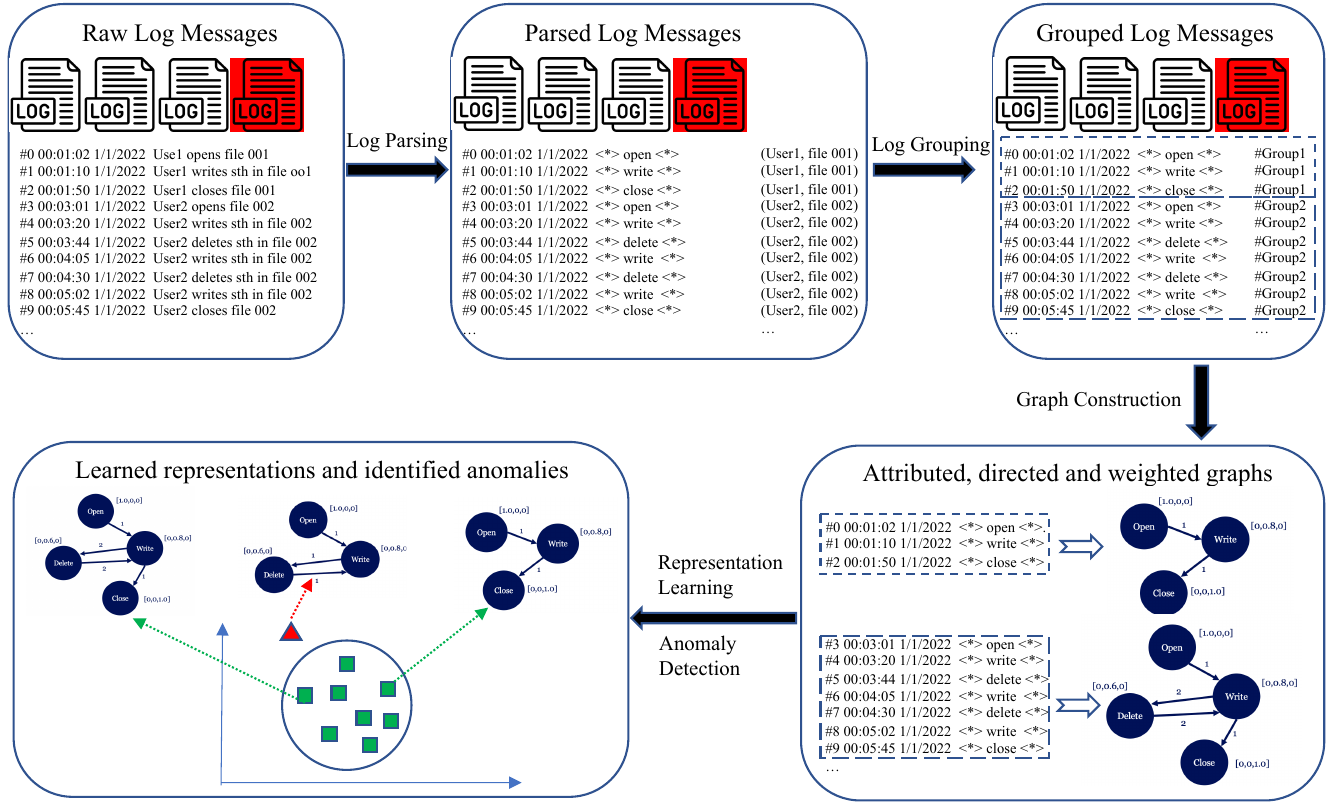}
	\caption{The Logs2Graphs pipeline. We use attributed, directed, and weighted graphs for representing the log files with high expressiveness, and integrate representation learning and anomaly detection for accurate anomaly detection. We use off-the-shelf methods for log parsing, log grouping, and graph construction.} \label{fig:Logs2Graphs}
\end{figure*}

After obtaining the multi-scale features $\{\textbf{Z}^{(0)},\textbf{Z}^{(1)},...,\textbf{Z}^{(k)}\}$, DiGCN defines an Inception block as 
\begin{equation}
	\textbf{Z} = \sigma \left(\Gamma \left(\textbf{Z}^{(0)},\textbf{Z}^{(1)},...,\textbf{Z}^{(k)}\right)\right),
\end{equation}
where $\sigma$ represents an activation function, and $\Gamma(\cdot)$ denotes a fusion operation, which can be summation, normalisation, and concatenation. In practice, we often adapt a fusion operation that keeps the output dimension unchanged, namely $\textbf{Z} \in \mathcal{R}^{|\mathcal{V}|\times f}$. As a result, the $i$-th row of $\textbf{Z}$ (namely $\textbf{Z}_{i}$) denotes the learned vector representation for node $v_{i}$ in a certain layer.

\section{Graph-Based Anomaly Detection for Event Logs}

We propose \textit{Logs2Graphs}, a graph-based log anomaly detection method tailored to event logs. The overall pipeline consists of the usual main steps, i.e., log parsing, log grouping, graph construction, graph representation learning, and anomaly detection, and is illustrated in Figure \ref{fig:Logs2Graphs}. Note that we couple the graph representation learning and anomaly detection steps to accomplish end-to-end learning once the graphs have constructed.

First, after collecting logs from a system, the \textit{log parsing} step extracts log events and log parameters from raw log messages. Since log parsing is not the primary focus of this article, we use Drain~\cite{he2017drain} for this task. Drain is a log parsing technique with fixed depth tree, and has been shown to generally outperform its competitors \cite{zhu2019tools}. We make the following assumptions on the log files:
\begin{itemize}
    \item Logs files are written in English;
    \item Each log message contains at least the following information: date, time, operation detail, and log identifier;
    \item The logs contain enough events to make the mined relationships (quantitative, sequential, structural) statistically meaningful, i.e., it must be possible to learn from the logs what the `normal' behaviour of the system is.
\end{itemize}

Second, the \textit{log grouping} step uses the log identifiers to divide the parsed log messages into log groups. Third, for each resulting group of log messages, the \textit{graph construction} steps builds an attributed, directed, and edge-weighted graph, as described in more detail in Subsection~5.1. Fourth and last, in an integrated step for \textit{graph representation learning and anomaly detection}, we learn a One-Class Digraph Inception Convolutional Network (OCDiGCN) based on the obtained set of log graphs. The resulting model can be used for graph-level anomaly detection. This model couples the graph representation learning objective and anomaly detection objective, and is thus trained in an end-to-end manner. The model, its training, and its use for graph-level anomaly detection are explained in detail in Subsection~5.2. %Accordingly, graphs with learned representation vectors far from the center are considered anomalous. We will elaborate on this later.

\subsection{Graph Construction}

\begin{figure}[h!]
\centering
\includegraphics[width=8cm]{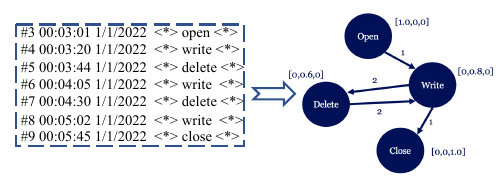}
\caption{The construction of an attributed, directed, and edge-weighted graph from a group of log messages.} \label{fig:GraphCons}
\end{figure}

We next explain how to construct an attributed, directed, and edge-weighted graph given a group of parsed log messages, and illustrate this in Figure~\ref{fig:GraphCons}. %Particularly, the motivation of graph construction is to keep everything relevant in log data.

First, we utilise nodes to represent different log events. As a result, the number of nodes depends on the number of unique log events that occur within the log group. Second, starting from the first line of log messages in chronological order, we add a directed edge from log event $L_{i}$ to $L_{j}$ and set its edge-weight to $1$ if the next event after $L_{i}$ is $L_{j}$. If the corresponding edge already exists, we increase its edge-weight by $1$. In this manner, we obtain a labelled, directed, and edge-weighted graph. 

However, using only the labels (e.g., \textit{open} or \textit{write}) of log events for graph construction may lead to missing important information. That is, we can  improve on this by explicitly taking the semantic information of log events into account, by which we mean that we should look at the text of the log event in entirety. Specifically, we generate a vector representation for each log event as follows:
\begin{enumerate}
    \item \textit{Preprocessing}: for each log event, we first remove non-character words and stop words, and split compound words into separate words;
    \item \textit{Word embedding}: we use Glove \cite{pennington2014glove}, a pre-trained word embedding model with 200 embedding dimensions to generate a vector representation for each word in a log event;
    \item \textit{Sentence embedding}: we generate a vector representation for each log event. Since the words in a sentence are usually not of equal importance, we use Term Frequency-Inverse Document frequency (TF-IDF) \cite{ramos2003using} to measure the importance of words. As a result, the weighted sum of word embedding vectors composes the vector representation of a log event.
\end{enumerate}
By augmenting the nodes with the vector representations of the log events as attributes, we obtain an attributed, directed, and edge-weighted graph.

\subsection{OCDiGCN: One-Class Digraph Inception Convolutional Nets}
%% Two things to add
%% 1. How to deal with the edge attributes E.
%% 2. Myabe we do not need the Inception Block since the original work is desinged for large graph. However, in our domain we only have small graphs. Preliminary results show the case that k=1 delivers the best reuslts. One possible reason is that our generated graph is usually small. If we set k>1, each node may receive Information for all nodes in a graph, resulting in the same representation for every node. 

We next describe One-Class Digraph Inception Convolutional Networks, abbreviated as OCDiGCN, a novel method for end-to-end graph-level anomaly detection. We chose to build on Digraph Inception Convolutional Networks (DiGCN) \cite{tong2020digraph} for their capability to handle directed graphs, which we argued previously is an advantage in graph-based log anomaly detection. 

Given that DiGCN was designed for node representation learning, we repurpose it for graph representation learning as follows:
\begin{equation}\label{equ:readout}
	\textbf{z} = \mathrm{Readout}(\textbf{Z}_{i} \mid i\in\{1,2,...,|\mathcal{V}|\}).
\end{equation}
That is, at the final iteration layer, we utilise a so-called $\mathrm{Readout}(\cdot)$ function to aggregate node vector representations to obtain a graph vector representation. Importantly, $\mathrm{Readout}(\cdot)$  can be a simple permutation-invariant function such as maximum, sum or mean, or a more advanced graph-level pooling function \cite{ying2018hierarchical}.

Next, note that DiGCN work did not explicitly enable learning edge features (i.e., $\textbf{Y}$). However, as DiGCN follows the Message Passing Neural Network (MPNN) framework \cite{gilmer2017neural}, incorporating $\textbf{Y}$ into Equation (1) and conducting computations in Equations (2-4) analogously enables learning edge features.

Now, given a set of graphs $\{\mathcal{G}_{1},...,\mathcal{G}_{m},...,\mathcal{G}_{M}\}$, we can use Equation (\ref{equ:readout}) to obtain an explicit vector representation for each graph, respectively. We denote the vector presentation of $\mathcal{G}_{m}$ learned by the DiGCN model as $\mathrm{DiGCN(\mathcal{G}_{m};\mathcal{H})}$. 

In graph anomaly detection, anomalies are typically identified based on a reconstruction or distance loss \cite{kim2022graph}. In particular, the One-Class Deep SVDD objective \cite{ruff2018deep} is commonly used for two reasons: it can  be easily combined with other neural networks, and more importantly, it generally achieves a state-of-the-art performance \cite{pang2021deep}. To detect anomalies, we thus train a one-class classifier by optimising the following One-Class Deep SVDD objective:
\begin{equation}\label{equ:OCDiGCN}
    \min\limits_{\mathcal{H}}\frac{1}{M}\sum_{m=1}^{M}\lVert\mathrm{DiGCN(\mathcal{G}_{m};\mathcal{H})-\textbf{o}}\rVert_{2}^{2}+\frac{\lambda}{2}\sum_{l=1}^{L}\lVert\textbf{H}^{(l)}\rVert_{F}^{2},
\end{equation}
where $\textbf{H}^{(l)}$ represents the trainable parameters of DiGCN at the $l$-th layer, namely $(\Theta^{(0)(l)},\Theta^{(1)(l)},...,\Theta^{(k)(l)})^{T}$,  $\mathcal{H}$ denotes $\{\textbf{H}^{(1)},...,\textbf{H}^{(L)}\}$,  $\lambda>0$ represents the weight-decay hyperparameter, $\lVert\cdot\rVert_{2}$ is the Euclidean norm, and $\lVert\cdot\rVert_{F}$ denotes the Frobenius norm. Moreover, $\textbf{o}$ is the center of the hypersphere in the learned representation space. Ruff et al.\ \cite{ruff2018deep} empirically found that setting $\textbf{o}$ to the average of the network representations (i.e., graph representations in our case) obtained by performing an initial forward pass is a good strategy. 

Ruff et al.\ \cite{ruff2018deep} also pointed out, however, that One-Class Deep SVDD classification may suffer from a hypersphere collapse, which will yield trivial solutions, namely mapping all graphs to a fixed center in the representation space. To avoid a hypersphere collapse, the hypersphere center $\textbf{o}$ is set to the average of the network representations, the bias terms in the neural networks are removed, and unbounded activation functions such as ReLU are preferred.

After training the model on a set of non-anomalous graphs (or with a very low proportion of anomalies), given a test graph $\mathcal{G}_{m}$, we define its distance to the center in the representation space as its anomaly score, namely
\begin{equation}
\label{equ:AnomalyScore}
    score(\mathcal{G}_{m}) = \lVert\mathrm{DiGCN(\mathcal{G}_{m};\mathcal{H})-\textbf{o}}\rVert_{2}.
\end{equation}

\begin{table}
  \caption{Summary of datasets. \#Events refers to the number of log event templates obtained using log parser Drain \cite{he2017drain}. \#Groups means the number of generated graphs. \#Anomalies represents the number of anomalous graphs. \#Nodes denotes the average number of nodes in generated graphs. \#Edges indicates the average number of edges in the generated graphs.}
  \label{tab:datasets}
  \begin{adjustbox}{width=\linewidth}
  \begin{tabular}{cccccc}
    \toprule
    Name&\#Events &\#Graphs &\#Anomalies &\#Nodes &\#Edges \\
    \midrule
    HDFS  & 48 &575,061 &16,838 & 7 & 20\\
    Hadoop & 683 & 978 & 811  & 34 & 120 \\
    BGL & 1848 &69,251 & 31,374  & 10 & 30 \\
    Spirit & 834 & 10,155 & 4,432  & 6 & 24\\
    Thunderbird & 1013 & 52,160 &6,814  & 16 & 52\\
  \bottomrule
\end{tabular}
\end{adjustbox}
\end{table}

\begin{table}
  \caption{Description of hyperparameters involved in OCDiGCN. \textbf{Range} indicates the values that we have tried on validation data, and boldfaced values are the values suggested to use in experiments. Particularly, for the embedding dimensions: 300 is suggested for BGL and 128 for others. For the batch sizes: 32 is suggested for HDFS and 128 for others. For the training epochs:  100 for BGL and Thunderbird, 200 for HDFS, 300 for Hadoop and 500 for Spirit are suggested.}
  \label{tab:hyperparameters}
  \centering
  \begin{adjustbox}{width=\linewidth}
  \begin{tabular}{ccc}
    \toprule
    \textbf{Symbol} & \textbf{Meaning} & \textbf{Range} \\
    \midrule
    $Bs$ & Batch size & \{16, 32, 64, \textbf{128}, 256, 512, 1024, 1536, 2048, 2560\}\\
    \hline
    $Op$ & optimisation method & Adam, \textbf{SGD}\\
    \hline
    $L$ & number of layers& $\{\textbf{1},2,3,4,5\}$\\
    \hline
    $\lambda$ & weight decay parameter & $\{\textbf{0.0001}, 0.001,0.01,0.1\}$ \\
    \hline
     $\eta$ & learning rate  & $\{0.0001,0.001,\textbf{0.01}\}$ \\
     \hline
     $k$ & proximity parameter & $\{\textbf{1},2\}$ \\
     \hline
     $\alpha$ & teleport probability  & $\{0.05, \textbf{0.1}, 0.2\}$ \\
     \hline
     $\Gamma$ & fusion operation if $k\geq2$ & sum, concatenation \\
     \hline
     $Re$ & readout function  & \textbf{mean}, sum, max \\
     \hline
     $d$ & embedding dimension & $\{32, 64, \textbf{128}, 256, 300\}$\\
     \hline
     $Ep$ & 
     Epochs for training & range(100,1000,50)\\
  \bottomrule
\end{tabular}
\end{adjustbox}
\end{table}

\textbf{Training and hyperparameters: } In summary, OCDiGCN is composed of an $L$-layer DiGCN architecture to learn node representations, plus a $\mathrm{Readout}(\cdot)$ function to obtain the graph representation. It is trained in an end-to-end manner via optimising the SVDD objective, which can be optimised using stochastic optimisation techniques such as Adam \cite{kingma2014adam}. Overall, OCDiGCN takes a collection of non-anomalous graphs and a set of hyperparameters, which are outlined in Table \ref{tab:hyperparameters}, as inputs. The pseudo-code for Logs2Graphs is given in Algorithm \ref{Alog:Logs2Graphs}.
\begin{algorithm}[h]
    \caption{Pseudo-code of Logs2Graphs}\label{Alog:Logs2Graphs}
\begin{flushleft}
\textbf{Input:} Training dataset $D_{tr}$, testing dataset $D_{ts}$, model $\theta$

\textbf{Output:} Predicted labels and explanations for $D_{ts}$
\end{flushleft}
    \begin{algorithmic}[1]
    \State $\hat{D}_{tr}, \hat{D}_{ts} \leftarrow Drain\_parse(D_{tr}), Drain\_parse(D_{ts})$
    %\State Parse $D_{tr}$ and $D_{ts}$ using Drain \cite{he2017drain} $\rightarrow$ Obtain parsed datasets $\hat{D}_{tr}$ and $\hat{D}_{ts}$
    \State Group $\hat{D}_{tr}$ and $\hat{D}_{ts}$ based on log identifier $\rightarrow$  Obtain grouped dataset $\tilde{D}_{tr}$ and $\tilde{D}_{ts}$ 
    \State Construct graphs using  $\tilde{D}_{tr}$ and $\tilde{D}_{ts}$ $\rightarrow$ Obtain graph sets $\textbf{Q}_{tr}$ and $\textbf{Q}_{ts}$
    \State Train the OCDiGCN model using Equation (\ref{equ:OCDiGCN}) with  $\textbf{Q}_{tr}$ $\rightarrow$ Obtain trained model $\hat{\theta}$
    \State Use $\hat{\theta}$ to predict anomalies in $\textbf{Q}_{ts}$ $\rightarrow$ Obtain a set of anomalies $\{Q_{1},...,Q_{n}\}$
    \State Generate explanations for $Q_{i} \in \{Q_{1},...,Q_{n}\}$
    \end{algorithmic}
\end{algorithm}

\subsection{Anomaly Explanation}
\label{Sec:AnomalyExplanation}
{\color{black} Our anomaly explanation method can be regarded as a decomposition method \cite{yuan2022explainability}, that is, we build a score decomposition rule to distribute the prediction anomaly score to the input space. Concretely, a graph $\mathcal{G}_{m}$ is identified as anomalous if and only if its graph-level representation has a large distance to the hyper-sphere center (Equation \ref{equ:AnomalyScore}). Further, the graph-level representation is obtained via a Readout$(\cdot)$ function applied on the node-level representations (Equation \ref{equ:readout}). Therefore, if the Readout$(\cdot)$ function is attributable (such as the sum or the mean), we can easily obtain the a small subset of important nodes (in the penultimate layer) whose node embeddings contribute the most to the distance. Specifically, the importance score of node $v_{j}$ (in the penultimate layer)  in a graph $\mathcal{G}_{m}$ is defined as
\begin{equation}
    \frac{|score(\mathcal{G}_{m}) - score(\mathcal{G}_{m}\setminus\{\textbf{Z}_{j}\})|}{ score(\mathcal{G}_{m})}
\end{equation}
where $score(\mathcal{G}_{m})$ is defined in Equation \ref{equ:AnomalyScore} and $score(\mathcal{G}_{m}\setminus\{\textbf{Z}_{j}\})$ is the anomaly score by removing the embedding vector of $v_{j}$ (namely $\textbf{Z}_{j}$) when applying the Readout function to obtain the graph-level representation.

Next, for each important node (with a high importance score) in the penultimate layer, we extend the LRP (Layerwise Relevance Propagation) algorithm \cite{bach2015pixel} to obtain a minor set of important nodes in the input layer (this is not the contribution of our paper and we simply follow the practice in \cite{schwarzenberg2019layerwise,baldassarre2019explainability}). If certain of these nodes are connected by edges, the resulting subgraphs can provide more meaningful explanations. As the LRP method generates explanations utilizing the hidden features and model weights directly, its explanation outcomes are deemed reliable and trustworthy \cite{li2022survey}.}

%Our future agenda involves the integration of GNN interpretability techniques \cite{yuan2022explainability} into Logs2Graphs. This integration aims to ascertain the importance of individual log events and subgraphs within identified anomalous graphs, which will offer valuable cues for subsequent root cause diagnosis.

\section{Experiments}

We perform extensive experiments to answer the following questions:
\begin{enumerate}
    \item \textbf{Detection accuracy:} How effective is \textit{Logs2Graphs} at identifying log anomalies when compared to state-of-the-art methods?
    %\item \textbf{End-to-end training vs. Two-stage training}: Is end-to-end training better than two-stage training in detecting log anomalies?
    \item \textbf{Directed vs. undirected graphs}: Is the directed log graph representation better than the undirected version for detecting log anomalies? 
    {\color{black} \item \textbf{Node Labels vs. Node Attributes}: How important is it to use semantic embeddings of log events as node attributes?}
    {\color{black}\item \textbf{Robustness analysis:} To what extent is Logs2Graphs robust to contamination of the training data?}
     {\color{black}\item \textbf{Ability to detect structural anomalies:} Can Logs2Graphs better capture structural anomalies and identify structurally equivalent normal instances than its contenders?}
    {\color{black}  \item \textbf{Explainability Analysis:} How understandable are the anomaly explanations given by Logs2Graphs?}
    \item \textbf{Sensitivity analysis:} How do the values of the hyperparameters influence the detection accuracy?
    \item \textbf{Runtime analysis:} What are the runtimes for the different methods?
    %\item \textbf{Model selection (Sensitivity Analysis):} Can the unsupervised model selection approach, namely ModelCentrality, select the best model? (Or a better-than-average model)
\end{enumerate}

\subsection{Experiment Setup}

\subsubsection{Datasets}
The five datasets that we use, summarised in Table~\ref{tab:datasets}, were chosen for three reasons: 1) they are commonly used for the evaluation of log anomaly detection methods; 2) they contain ground truth labels that can be used to calculate evaluation metrics; and 3) they include log identifiers that can be used for partitioning log messages into groups.  For each group of log messages in a dataset, we label the group as anomalous if it contains at least one anomaly.
More details are given as follows:
\begin{itemize}
    \item HDFS \cite{xu2009detecting} consists of Hadoop Distributed File System logs obtained by running 200 Amazon EC2 nodes. These logs contain \textit{block\_id}, which  can be used to group log events into different groups. Moreover, these logs are manually labeled by Hadoop experts.
    \item Hadoop \cite{lin2016log} was collected from a Hadoop cluster consisting of 46 cores over 5 machines. The \textit{ContainerID} variable is used to divide log messages into different groups. 
    \item BGL, Spirit, and Thunderbird contain system logs collected from the BlueGene/L (BGL),
    Spirit, and Thunderbird supercomputing systems located at Sandia National Labs, respectively. For those datasets, each log message was manually inspected by engineers and labelled as normal or anomalous. For BGL, we use all log messages, and group log messages based on the \textit{Node} variable. For Spirit and Thunderbird, we only use the first 1 million and first 5 million log messages for evaluation, respectively. Furthermore, for these two datasets, the \textit{User} is used as log identifier to group log messages. However, considering that an ordinary user may generate hundreds of thousands of logs, we regard every 100 consecutive logs of each user as a group. If the number of logs is less than 100, we also consider it as a group.
\end{itemize}

\subsubsection{Baselines}
To investigate the performance of $Logs2Graphs$, we compare it with the following seven log anomaly detection methods: Principal Component Analysis (PCA) \cite{xu2009detecting}, One-Class SVM (OCSVM) \cite{scholkopf2001estimating}, Isolation Forest (iForest) \cite{liu2008isolation},  HBOS \cite{goldstein2012histogram}, DeepLog \cite{du2017deeplog}, LogAnomaly \cite{meng2019loganomaly}, and AutoEncoder \cite{farzad2020unsupervised}, and one state-of-the-art graph level anomaly detection method: GLAM \cite{zhao2022graph}.

We choose these methods as baselines because they are often regarded to be representatives of traditional machine learning-based (PCA, OCSVM, IForest, HBOS) and deep learning-based approaches (DeepLog, LogAnomaly and AutoEncoder), respectively. All methods are unsupervised or semi-supervised methods that do not require labelled anomalous samples for training the models.
%Except for RobustLog, the remaining methods are unsupervised or semi-supervised methods that do not require labelled anomalous samples to train the model. %Although Logs2Graphs is unsupervised, we intend to compare it with RobustLog, aiming to investigate its performance gap with supervised methods. Besides, despite claiming to be a semi-supervised method, DeepLog and LogAnomaly require a small set of labeled data (both anomalous and normal samples) as validation dataset for hyperparameter tuning.

\subsubsection{Evaluation Metrics} The Area Under Receiver Operating Characteristics Curve (ROC AUC) and the Area Under the Precision-Recall Curve (PRC AUC) are widely used to quantify the detection accuracy of anomaly detection \cite{aggarwal2017introduction}. This is mainly because they can provide a single value that summarizes the overall performance of the anomaly detection model across various thresholds. In contrast, other metrics such as Precision, Recall and F1-score depend on choosing a threshold to determine whether an instance is anomalous or normal. Consequently, different thresholds can result in different values. Therefore, we employ ROC AUC and PRC AUC to evaluate and compare the different log anomaly detection methods. PRC AUC is also known as Average Precision (AP). For both ROC AUC and PRC AUC, values closer to 1 indicate better performance.

\subsection{Model Implementation and Configuration}

Traditional machine learning based approaches---such as PCA, OCSVM, iForest, and HBOS---usually first transform logs into log event count vectors, and then apply traditional anomaly detection techniques to identify anomalies. For these methods, we utilise their open-source implementations provided in PyOD \cite{zhao2019pyod}. Meanwhile, for deep learning methods DeepLog, LogAnomaly, and AutoEncoder, we use their open-source implementations in Deep-Loglizer \cite{chen2021experience}. For these  methods, we use their default hyperparameter values.

For all deep learning based methods, the experimental design adopted in this study
follows a train/validation/test strategy with a distribution of $70\%:5\%:25\%$ for normal instances. Specifically, the model was trained using $70\%$ of normal instances, while $5\%$ of normal instances and an equal number of abnormal instances were employed for validation (i.e., hyperparameter tuning). The remaining $25\%$ of normal instances and the remaining abnormal instances were used for testing. Table~\ref{tab:hyperparameters} summarises the hyperparameters involved in OCDiGCN as well as their recommended values.

%The sensitivity analysis on OCDiGCN indicates that it does  not need hyperparameter tuning, as it performed well across a wide range of values. Therefore, for OCDiGCN the ratio of validation instances was reduced to $1\%$ of normal instances and an equal number of abnormal instances (leaving more instances for training). The hyperparameter values selected using this reduced validation set were found to be almost identical to those selected using the larger set.

We implemented and ran all algorithms in Python 3.8 (using PyTorch \cite{paszke2019pytorch} and PyTorch Geometric \cite{fey2019fast} libraries when applicable), %on a computer with Apple M1 chip 8-core CPU and 16GB unified memory. 
on a workstation equipped with an Intel i7-11700KF CPU and Nvidia RTX3070 GPU.
%For reproducibility, all code and datasets will be released on GitHub. 
For reproducibility, all code and datasets are released on GitHub.\footnote{https://github.com/ZhongLIFR/Logs2Graph}

\begin{table*}
  \caption{Anomaly detection accuracy on five benchmark datasets for \textit{Logs2Graphs} and its eight competitors. AP and RC denote Average Precision and ROC AUC, respectively. HDFS, BGL, and Thunderbird have been downsampled to 10,000 graphs each while maintaining the original anomaly rates. For each method on each dataset, to mitigate potential biases arising from randomness, we conducted ten experimental runs with varying random seeds and report the average values along with standard deviations of AP and RC. Moreover, we highlight the best results with \textbf{bold} and the runner-up with \underline{underline}. %To conserve spaces, we use the following abbreviations: PCA (M1), OCSVM (M2), iForest (M3), HBOS (M4), DeepLog (M5), LogAnomaly (M6), AutoEncoder (M7), GLAM (M8) and Logs2Graphs (M).
  }
  \label{tab:results1}
  \begin{adjustbox}{width=\linewidth}
  \begin{tabular}{ccccccccccc}
    \hline
      &\multicolumn{2}{c}{\textbf{HDFS}}&\multicolumn{2}{c}{\textbf{Hadoop}}&\multicolumn{2}{c}{\textbf{BGL}}&\multicolumn{2}{c}{\textbf{Spirit}} &\multicolumn{2}{c}{\textbf{Thunderbird}}\\
    \cline{2-11}
    Method &AP &RC &AP &RC &AP &RC &AP &RC &AP &RC \\
    \midrule
   PCA 
   & \underline{0.91}$\pm$0.03
   & \textbf{1.0}$\pm$0.00  
   & 0.84$\pm$0.00  
   & 0.52$\pm$0.00 
   & 0.73$\pm$0.01 
   & 0.82$\pm$0.00 
   & 0.31$\pm$0.00 
   & 0.19$\pm$0.00 
   & 0.11$\pm$0.00 
   & 0.34$\pm$0.01
   \\
    OCSVM 
    &0.18$\pm$0.01 
    & 0.88$\pm$0.01 
    & 0.83$\pm$0.00   
    & 0.45$\pm$0.00   
    & 0.47$\pm$0.00  
    & 0.47$\pm$0.01  
    & 0.34$\pm$0.00  
    & 0.29$\pm$0.00  
    & 0.12$\pm$0.00  
    & 0.45$\pm$0.01 
    \\
    IForest
    &0.73$\pm$0.04 
    & 0.97$\pm$0.01 
    & 0.85$\pm$0.01   
    & 0.55$\pm$0.01   
    & 0.79$\pm$0.01 
    & 0.83$\pm$0.01 
    & 0.32$\pm$0.03  
    & 0.23$\pm$0.02  
    & 0.11$\pm$0.01  
    & 0.24$\pm$0.10 
    \\
    HBOS 
    &0.74$\pm$0.04 
    &\underline{0.99}$\pm$0.00 
    &0.84$\pm$0.00   
    &0.50$\pm$0.00   
    &0.84$\pm$0.02   
    &0.87$\pm$0.03   
    &0.35$\pm$0.00  
    &0.22$\pm$0.00  
    &0.15$\pm$0.01  
    &0.29$\pm$0.05 
    \\
    \hline
    DeepLog 
    & \textbf{0.92}$\pm$0.07 
    & 0.97$\pm$0.04 
    & \textbf{0.96}$\pm$0.00 
    & 0.47$\pm$0.00 
    &0.89$\pm$0.00 
    &0.72$\pm$0.00 
    &\underline{0.99}$\pm$0.00  
    &\underline{0.97}$\pm$0.00  
    &\underline{0.91}$\pm$0.01  
    & \underline{0.96}$\pm$0.00 
    \\
    LogAnomaly 
    &0.89$\pm$0.09 
    & 0.95$\pm$0.05
    & \textbf{0.96}$\pm$0.00 
    &0.47$\pm$0.00 
    &0.89$\pm$0.00 
    &0.72$\pm$0.00 
    &\underline{0.99}$\pm$0.00  
    &\underline{0.97}$\pm$0.00  
    &0.90$\pm$0.01  
    &\underline{0.96}$\pm$0.00 
    \\
    AutoEncoder 
    &0.71$\pm$0.03 
    & 0.84$\pm$0.01 
    &\textbf{0.96}$\pm$0.00 
    &0.52$\pm$0.00 
    &0.91$\pm$0.01  
    &0.79$\pm$0.02  
    &0.96$\pm$0.00  
    &0.92$\pm$0.01  
    &0.44$\pm$0.02 
    &0.46$\pm$0.05
    \\
    \hline
    GLAM 
    &0.78$\pm$0.08 
    &0.89$\pm$0.04
    & \underline{0.95}$\pm$0.00 
    & \textbf{0.61}$\pm$0.00 
    &\underline{0.94}$\pm$0.02 
    &\underline{0.90}$\pm$0.03 
    &0.93$\pm$0.00 
    &0.91$\pm$0.00 
    &0.75$\pm$0.02
    &0.85$\pm$0.01
    \\
    %HDFS: epoch = 200, LR=0.01, hid_dim =128
    %Hadoop: epoch = 150, LR=0.01, hid_dim =128
    %BGL: epoch = 100, LR=0.01, hid_dim =300
    %Spirit: epoch = 2000, LR=0.01, hid_dim =128
    %Thunderbird: epoch = 100, LR=0.01, hid_dim =128
    Logs2Graphs 
    &0.87$\pm$0.04 
    &0.91$\pm$0.02 
    & \underline{0.95}$\pm$0.00 
    & \underline{0.59}$\pm$0.00 
    & \textbf{0.96}$\pm$0.01 
    &\textbf{0.93}$\pm$0.01 
    & \textbf{1.0}$\pm$0.00  
    & \textbf{1.0}$\pm$0.00  
    & \textbf{0.99}$\pm$0.00 
    & \textbf{1.0}$\pm$0.00
    \\
  \bottomrule
\end{tabular}
\end{adjustbox}

%%1. LSTM needs of a lot of training time
%%2. LSTM needs validation dataset
%%3. LSTM performs poorly using session windows (hard to train)
%%4. the inplementation of LogAnomaly in "LogDeep" is not reliable, we have to use its version in "DeepLoglizer".
%%5. Adding AutoEncoder and Transformer (this one maybe too slow) in "DeepLoglizer".
%%6. LSTM may need much more training data (test it using different amount of training dataset)
%%7. Test Logs2Graphs using sliding window, maybe it performs better.
%%8. Autoencoder needs to specify the percentage of anomalies for prediction in test dataset.

%\footnotesize{$^a$ $ $Both DeepLog and LogAnomaly perform poorly if session windows are used. Instead, they perform substantially better if time windows are used. Therefore, we report the latter case for them.}
\end{table*}

\subsection{Comparison to the state of the art}
We first compare $Logs2Graphs$ to the state of the art. We have the following main observations according to the results in Table~\ref{tab:results1}:
\begin{itemize}
    \item In terms of ROC AUC, \textit{Logs2Graphs} achieves the best performance against its competitors on three out of five datasets. Particularly, \textit{Logs2Graphs} outperforms the closet competitor on BGL with 9.6\% and delivers remarkable results (i.e., an ROC AUC larger than 0.99) on Spirit and Thunderbird. Similar observations can be made for Average Precision.
    \item Deep learning based methods generally outperform the traditional machine learning based methods. One possible reason is that traditional machine learning based methods only leverage log event count vectors as input, which makes them unable to capture and exploit sequential relationships between log events and the semantics of the log templates. 
    \item The performance of (not-graph-based) deep learning methods is often inferior to that of \textit{Log2Graphs} on the more complex datasets, i.e., BGL, Spirit, and Thunderbird, which all contain hundreds or even thousands of log templates. This suggests that LSTM-based models may not be well suited for logs with a large number of log templates. One possible reason is that the test dataset contains many unprecedented log templates, namely log templates that are not present in the training dataset.
    \item In terms of ROC AUC score, all methods except for OCSVM and AutoEncoder achieve impressive results (with $RC> 0.91$) on HDFS. One possible reason is that HDFS is a relatively simple log dataset that contains only 48 log templates. Concerning AP, PCA and LSTM-based DeepLog achieve impressive results (with $AP>0.89$) on HDFS. Meanwhile, \textit{Logs2Graphs} obtains a competitive performance (with $AP=0.87$) on HDFS.
    %For BGL, on average 664 new templates show while testing.
    %For HDFS, pn average 15 new templates show while testing.
    
    %\item In terms of Average Precision score, \textit{Logs2Graphs} achieves the best performance against its competitors on 3 out of 5 datasets. Particularly, \textit{Logs2Graphs} outperforms the closet competitor on BGL with 8.4\% and delivers remarkable results (namely large than 0.990) on Spirit and Thunderbird.
\end{itemize}

\subsection{Directed vs. undirected graphs} 

To investigate the practical added value of using \textit{directed} log graphs as opposed to \textit{undirected} log graphs, we convert the logs to attributed, undirected, and edge-weighted graphs, and apply GLAM \cite{zhao2022graph}, a graph-level anomaly detection method for undirected graphs. We use the same graph construction method as for \textit{Logs2Graphs}, except that we use undirected edges. Similar to our method, GLAM also couples the graph representation learning and anomaly detection objectives by optimising a single SVDD objective. The key difference with OCDiGCN is that GLAM leverages GIN \cite{xu2018powerful}, which can only tackle undirected graphs, while OCDiGCN utilises DiGCN \cite{tong2020digraph} that is especially designed for directed graphs. 

The results in Table \ref{tab:results1} indicate that GLAM's detection performance is comparable to that of most competitors. However, it consistently underperforms on all datasets, except for Hadoop, when compared to \textit{Logs2Graphs}. Given that the directed vs undirected representation of the log graphs is the key difference between the methods, a plausible explanation is that directed graphs have the capability to retain the temporal sequencing of log events, whereas undirected graphs lack this ability. Consequently, GLAM may encounter difficulties in detecting sequential anomalies and is outperformed by \textit{Logs2Graphs}.

\subsection{Node Labels vs. Node Attributes} 
\begin{figure}[h!]
\centering
\includegraphics[width=8cm]{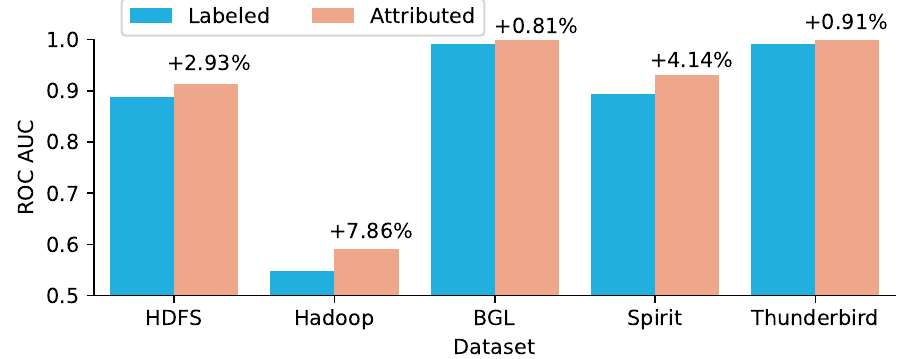}
\caption{The comparative performance analysis of Logs2Graphs, measured by ROC AUC, demonstrating the distinction between utilizing node semantic attributes and node labels.} \label{fig:Attributed}
\end{figure}

{\color{black}To investigate the importance of using semantic embeddings of log events as node attributes, we replace the node semantic attributes with one-hot-encoding of node labels (i.e., using an integer to represent a log event). The performance comparisons in terms of ROC AUC for Logs2Graphs are depicted in Figure \ref{fig:Attributed}, which shows that using semantic embeddings is always superior to using node labels. Particularly, it can lead to a substantial performance improvement on the Hadoop, Spirit and HDFS datasets. The PRC AUC results show a similar behaviour and thus are omitted.}

\subsection{Robustness to Contamination}
\begin{figure}[h!]
\centering
\includegraphics[width=8cm]{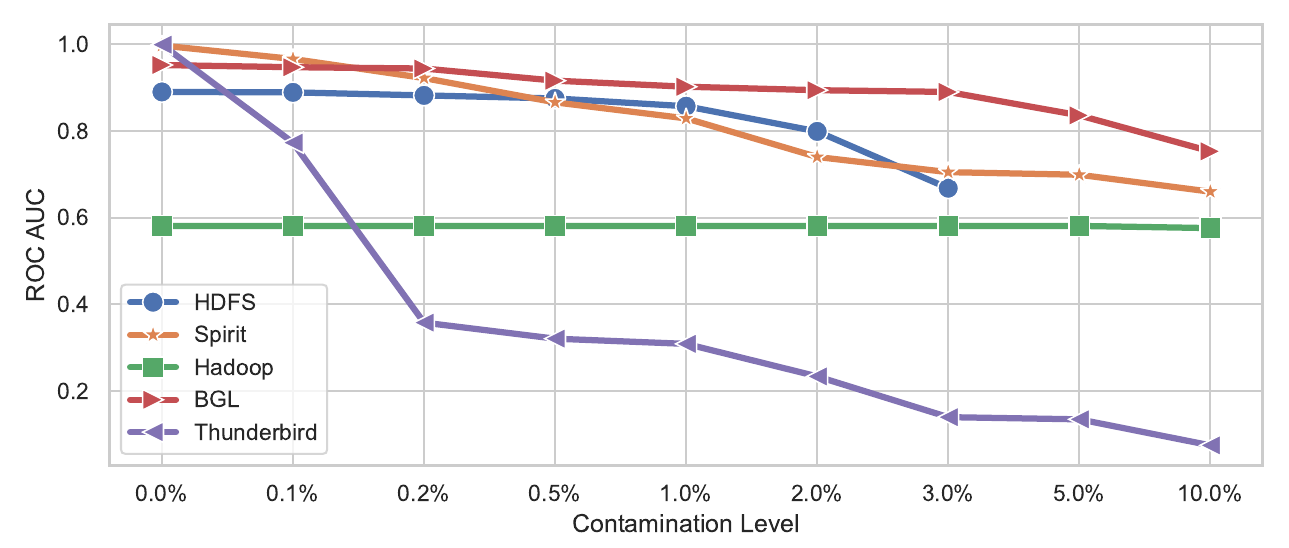}
\caption{ROC AUC results of Logs2Graphs w.r.t. a wide range of contamination levels. Results are averaged over 10 runs. Particularly, HDFS contains only 3\% anomalies and thus results at 5\% and 10\% are not available. } \label{fig:Contam}
\end{figure}

To investigate the robustness of Logs2Graphs when the training dataset is contaminated (namely integrating anomalous graphs in training data), we report its performance in terms of ROC AUC under a wide range of contamination levels. Figure \ref{fig:Contam} shows that the performance of Logs2Graphs decreases with an increase of contamination in the training data. The PRC AUC results show a similar behaviour and thus are omitted. Hence, it is important to ensure that the training data contains only normal graphs (or with a very low proportion of anomalies).

\subsection{Ability to Detect Structural Anomalies and Recognise Unseen Normal Instances}
{\color{black}
To showcase the effectiveness of different neural networks in detecting structural anomalies, we synthetically generate normal and anomalous directed graphs as shown in Figure \ref{fig:Synthetic}. As Deeplog, LogAnomaly and AutoEncoder require log sequences as inputs, we convert directed graphs into sequences by sequentially presenting the endpoints pair of each edge. Moreover, for GLAM we convert directed graphs into undirected graphs by turning each directed edge into an undirected edge. 

\begin{figure}[h!]
\centering
\includegraphics[width=8cm]{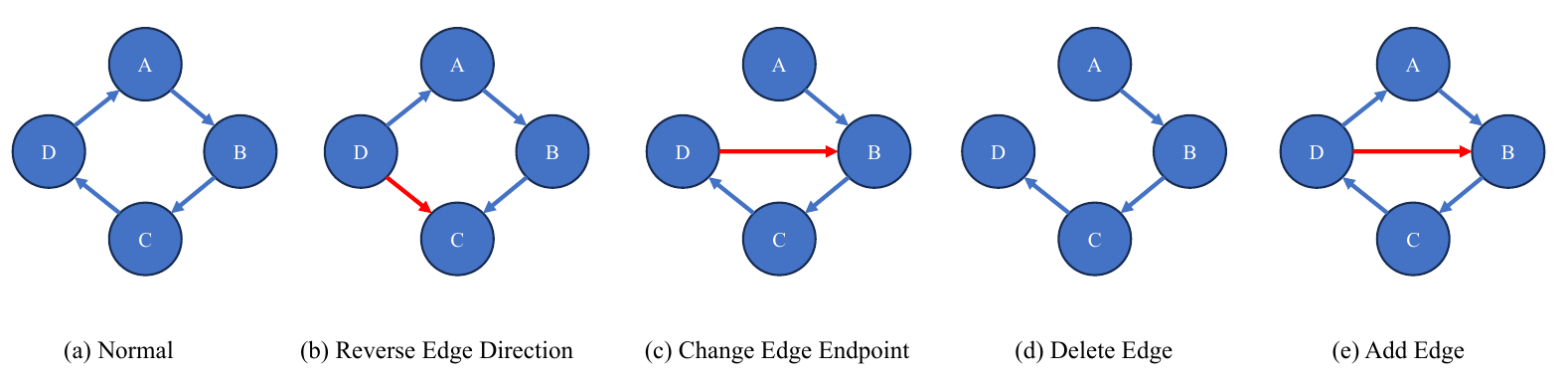}
\caption{Synthetic generation of normal (10000) and structurally anomalous (200 each) graphs.} \label{fig:Synthetic}
\end{figure}

Moreover, to investigate their capability of recognising unseen but structurally equivalent normal instances, we generate the following normal log sequences based on the synthetic normal graph as training data: $A \rightarrow B \rightarrow C \rightarrow D \rightarrow A$ (1000), $B \rightarrow C \rightarrow D \rightarrow A \rightarrow B$ (1000) and $C \rightarrow D \rightarrow A \rightarrow B \rightarrow C$ (1000), and the following as test dataset: $D \rightarrow A \rightarrow B \rightarrow C \rightarrow D$ (1000).

The results in Table \ref{tab:StructuralAnomalies} indicate that Logs2Graphs, Deeplog and LogAnomaly can effectively detect structural anomalies, while AutoEncoder and GLAM fail in some cases. However, log sequences based methods, namely Deeplog, LogAnomaly and AutoEncoder, can lead to high false positive rates due to their inability of recognising unseen but structurally equivalent normal instances.
}
\begin{table*}
  \caption{ROC AUC results (higher is better) of detecting structural anomalies and False Positive Rate (lower is better) of recognising unseen normal instances. S1: Reverse Edge Direction; S2: Change Edge Endpoint; S3: Delete Edge; S4: Add Edge; N1: Unseen normal instances.}
  \label{tab:StructuralAnomalies}
  \centering
  \begin{tabular}{cccccc}
    \toprule
    \textbf{Case} & Deeplog & LogAnomaly & AutoEncoder &GLAM &Ours  \\
    \midrule
    S1 (ROC) &1.0 &1.0 &0.0 &0.0 &1.0\\
    S2 (ROC) &1.0 &1.0 &0.50 &1.0 &1.0\\
    S3 (ROC) &1.0 &1.0 &1.0 &1.0 &1.0\\
    S4 (ROC) &1.0 &1.0 &1.0 &1.0 &1.0\\
    \hline
    N1 (FPR) &100\% &100\% &100\% &0\%  &0\%\\
  \bottomrule
\end{tabular}
\end{table*}

\subsection{Anomaly Explanation} 
\begin{figure}[h!]
\centering
\includegraphics[width=8cm]{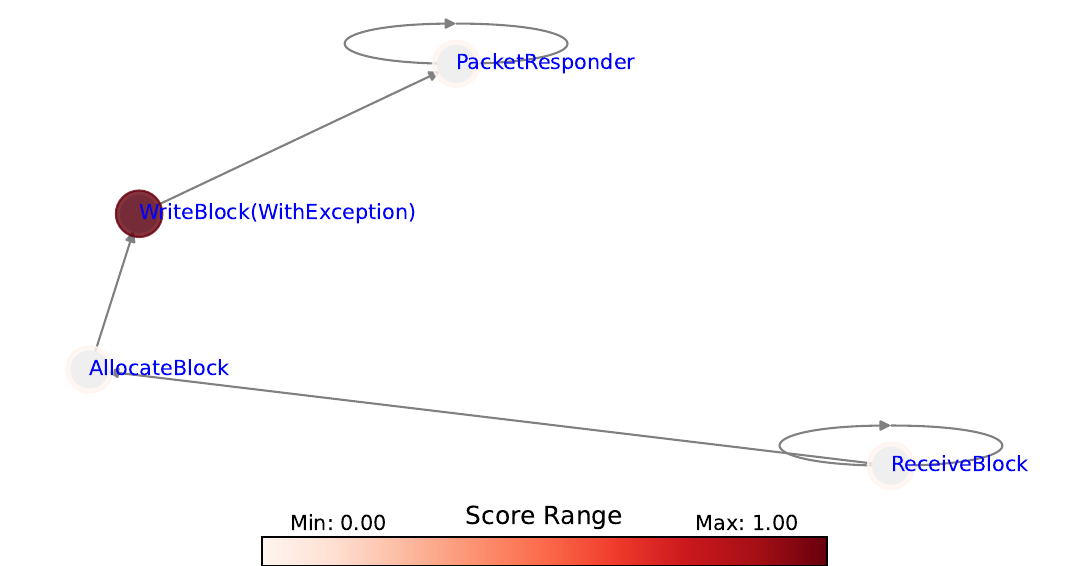}
\caption{Example of anomaly explanation with HDFS (the log event templates are simplified for better visualisation).} \label{fig:AnomalyExplanation}
\end{figure}

{\color{black}
Figure \ref{fig:AnomalyExplanation} provides an example of log anomaly explanation with the HDFS dataset. For each detected anomalous log graph (namely a group of logs), we first quantify the importance of nodes according to the description in Section \ref{Sec:AnomalyExplanation}. Next, we visualise the anomalous graph by assigning darker shade of red to more important nodes. In this example, the node ``WriteBlock(WithException)'' contributes the most to the anomaly score of an anomalous log group and thus is highlighted in red.}

\subsection{Sensitivity Analysis} 
%\begin{figure}%
    %\centering
    %\subfloat[\centering ]{{\includegraphics[width=7cm]{number_layers.pdf}}}%
    %\qquad
    %\subfloat[\centering ]{{\includegraphics[width=4.5cm]{runtime_spirit.pdf}}}%
    %\caption{(a) Sensitivity analysis with respect to the number of layers on five datasets; (b) Comparisons of runtime on Spirit dataset, where we only report the training time per epoch for neural networks based methods. Experiments are repeated 10 times.}%
    %\label{fig:Analysis}%
%\end{figure}

\begin{figure}[tbh]
\centering
\includegraphics[width=8cm]{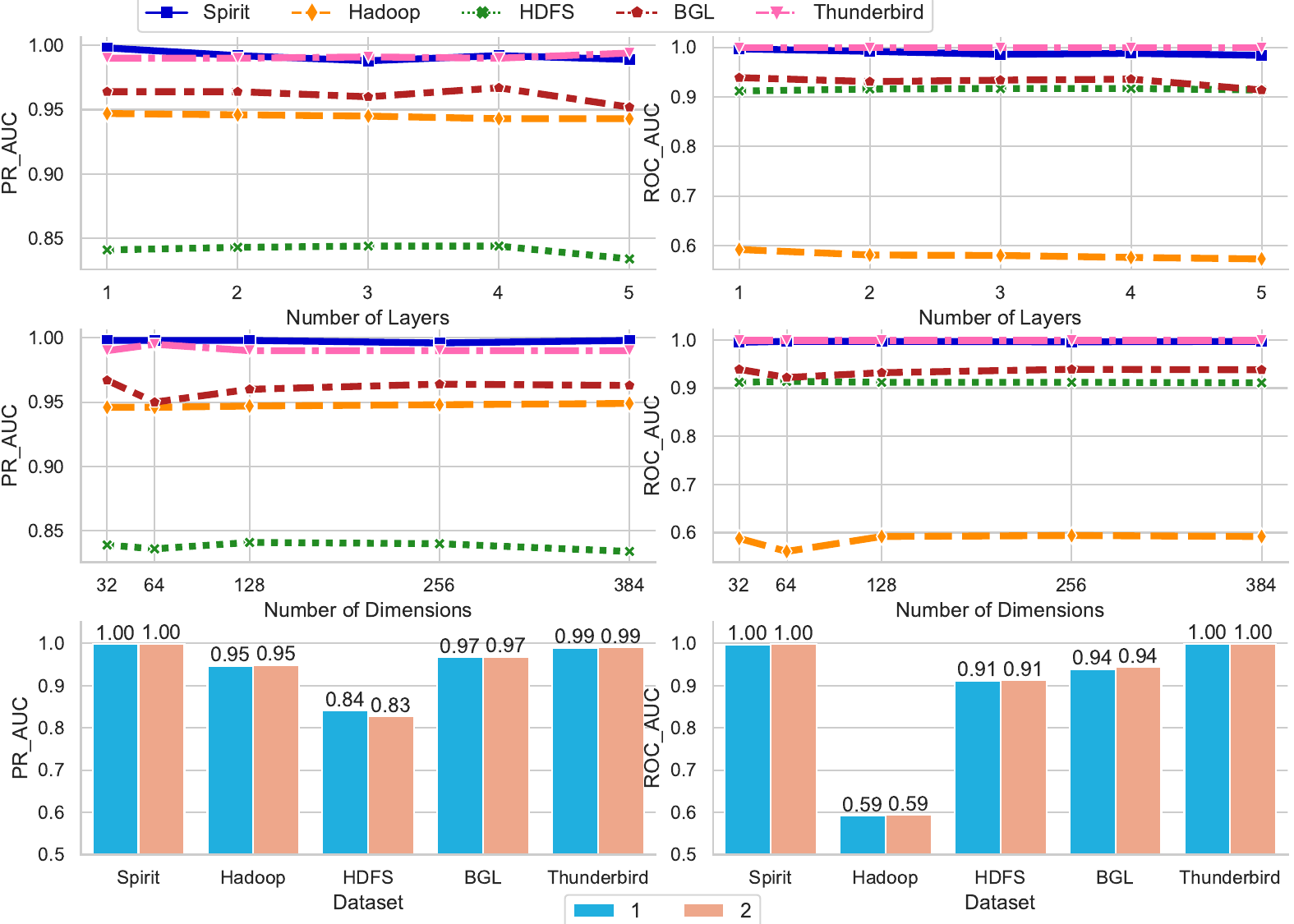}
\caption{The effects of the number of layers (top row), the embedding dimensions (middle row) and the proximity parameter (bottom row) on AP (left column) and ROC AUC (right column).} \label{fig:SA}
\end{figure}

We examine the effects of three hyperparameters in OCDiGCN on the detection performance.

\textbf{The Number of Convolutional Layers:} $L$ is a potentially important parameter as it determines how many convolutional layers to use in OCDiGCN. Figure \ref{fig:SA} (top row) depicts PRC AUC and ROC AUC for the five benchmark datasets when $L$ is varied from 1 to 5. We found that $L=1$ yields consistently good performance. As the value of $L$ is increased, there is only a slight enhancement in the resulting performance or even degradation, while the associated computational burden increases substantially. We thus recommend to set $L=1$.

\textbf{The Embedding Dimension $d$:} From Table \ref{fig:SA} (middle row) , one can see that $d=128$ yields good performance on Spirit, Hadoop, HDFS and Thunderbird, while further increasing $d$ obtains negligible performance improvement or even degradation. However, an increase of $d$ on BGL leads to substantially better performance. One possible reason is that BGL is a complex dataset wherein anomalies and normal instances are not easily separable on lower dimensions.

\textbf{The Proximity Parameter $k$:}
As this parameter increases, a node can gain more information from its further neighbours. Figure \ref{fig:SA} (bottom row) contrasts the detection performance when $k$ is set to 1 and 2, respectively. Particularly, we construct one Inception Block when $k = 2$, using concatenation to fuse the results.

We observe that there is no large improvement in performance when using a value of $k=2$ in comparison to $k=1$. It is important to recognize that a node exhibits 0th-order proximity with itself and 1st-order proximity with its immediately connected neighbors. If $k=2$, a node can directly aggregate information from its 2nd-order neighbours. As described in Table 1, graphs generated from logs usually contain a limited number of nodes, varying to 6 to 34. Therefore, there is no need to utilise the Inception Block, which was originally designed to handle large graphs in \cite{tong2020digraph}.

\subsection{Runtime Analysis}
Traditional machine learning methods, including PCA, OCSVM, IForest and HBOS, usually perform log anomaly detection in a transductive way. In other words, they require the complete dataset beforehand and do not follow a train-and-test strategy. In contrast, neural network based methods, such as DeepLog, LogAnomaly, AutoEncoder, and Logs2Graphs, perform log anomaly detection in an inductive manner, namely following a train-and-test strategy. 

\begin{figure}[h!]
\centering
\includegraphics[width=8cm]{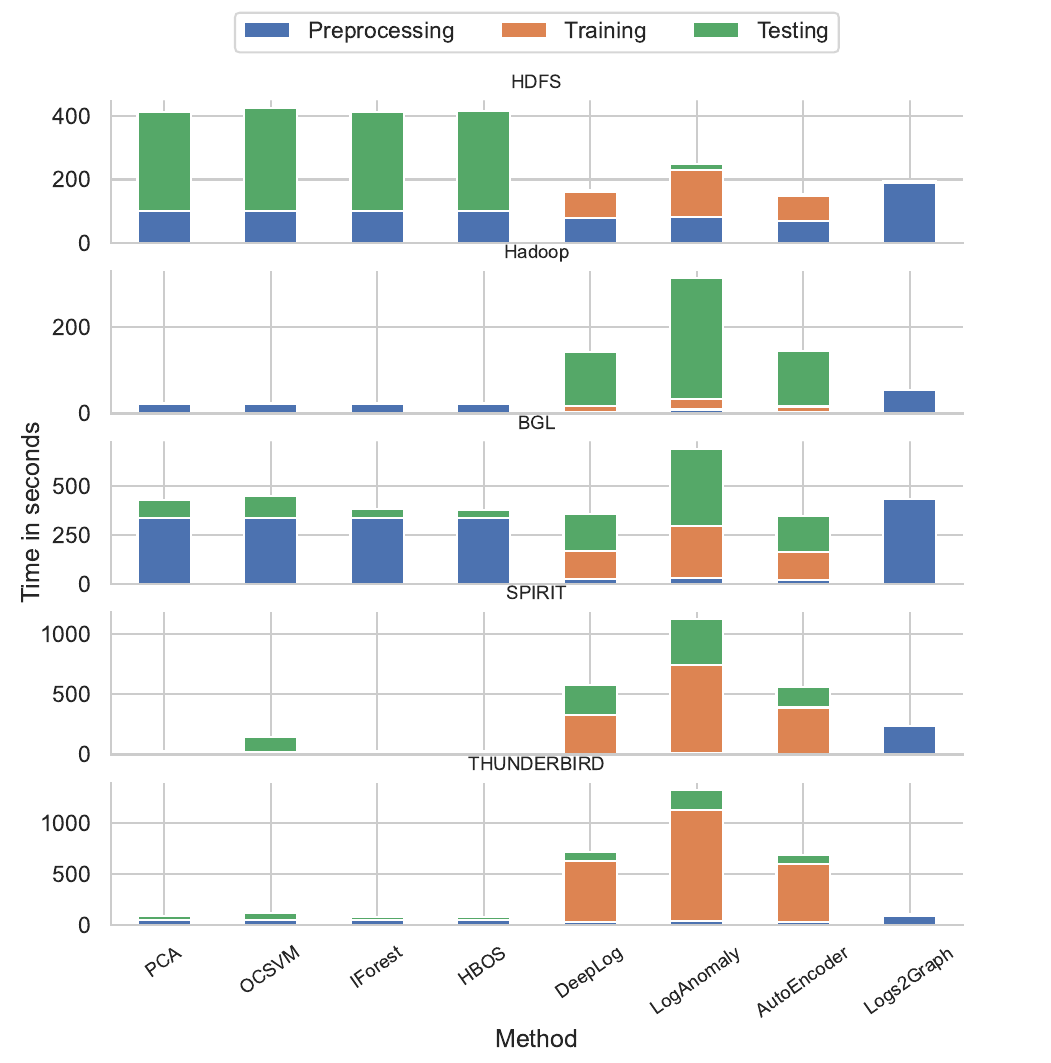}
\caption{Runtime for all eight methods on all datasets, wherein HDFS, BGL, and Thunderbird have been downsampled to 10,000 graphs. Runetimes are averaged over 10 repetitions. We report the training time per epoch for neural network based methods.} \label{fig:TCA}
\end{figure}

Figure \ref{fig:TCA} shows that most computational time demanded by \textit{Logs2Graphs} is allocated towards the graph generation phase. In contrast, the training and testing phases require a minimal time budget. The graph generation phase can be amenable to parallelisation though, thereby potentially reducing the overall processing time. As a result, \textit{Logs2Graphs} shows great promise in performing online log anomaly detection. Meanwhile, other neural networks based models---such as DeepLog, LogAnomaly, and AutoEncoder---demand considerably more time for the training and testing phases.

%%Things to do for this paper:
%1. Inception modular is not useful for us
%2. Check the implementation again
%3. Hyperparameter tunning (Model Selection)
%4. Run experiments on larger datasets?
%5. How DiGCN deal with edge-weights matrix (combined with code to explain it)
%6. Using sliding windows to test our method also (Maybe it performs better) -> It is slower and obtains worse results.

\section{Threats to Validity}
{\color{black}
We have discerned several factors that may pose a threat to the validity of our findings.

\textbf{Limited Datasets.} Our experimental protocol entails utilizing five publicly available log datasets, which have been commonly employed in prior research on log-based anomaly detection. However, it is important to acknowledge that these datasets may not fully encapsulate the entirety of log data characteristics. To address this limitation, our future work will conduct experiments on additional datasets, particularly those derived from industrial settings, in order to encompass a broader range of real-world scenarios.

\textbf{Limited Competitors.} This study focuses solely on the experimental evaluation of eight competing models, which are considered representative and possess publicly accessible source code. However, it is worth noting that certain models such as GLAD-PAW did not disclose their source code and it requires non-trivial efforts to re-implement these models. Moreover, certain models such as CODEtect require several months to conduct the experiments on our limited computing resources. For these  reasons, we exclude them from our present evaluation. In subsequent endeavors, we intend to re-implement certain models and attain more computing resources to test more models.

\textbf{Purity of Training Data.} The purity of training data is usually hard to guarantee in practical scenarios. Although Logs2Graphs is shown to be robust to very small contamination in the training data, it is critical to improve the model robustness by using techniques such as adversarial training \cite{bai2021recent} in the future. 

\textbf{Graph Construction.} The graph construction process, especially regarding the establishment of edges and assigning edge weights, adheres to a rule based on connecting consecutive log events. However, this rule may be considered overly simplistic in certain scenarios. Therefore, application-specific  techniques will be explored to construct graphs in the future.
}
%The graph construction, especially in the part concerning edges and edge weights, follows a rule (linking pairs of consecutive log events) which in some scenarios could be simplistic.

\section{Conclusions}

We introduced \textit{Logs2Graphs}, a new approach for unsupervised log anomaly detection. It first converts log files to attributed, directed, and edge-weighted graphs, translating the problem to an instance of graph-level anomaly detection. Next, this problem is solved by OCDiGCN, a novel method based on graph neural networks that performs graph representation learning and graph-level anomaly detection in an end-to-end manner. Important properties of OCDiGCN include that it can deal with directed graphs and do unsupervised learning.

Extensive results on five benchmark datasets reveal that \textit{Logs2Graphs} is at least comparable to and often outperforms state-of-the-art log anomaly detection methods such as DeepLog and LogAnomaly. Furthermore, a comparison to a similar method for graph-level anomaly detection on \textit{undirected} graphs demonstrates that using directed log graphs leads to better detection accuracy in practice. 

\section*{CRediT authorship contribution statement}
\textbf{Zhong Li}: Conceptualization, Methodology, Validation, Investigation, Software, Writing - original draft,
Visualisation, Project Administration. \textbf{Jiayang Shi}: Methodology, Validation, Software, Writing - original draft.
\textbf{Matthijs van Leeuwen}: Methodology, Validation, Writing - review \& editing, Funding acquisition.

\section*{Declaration of competing interest}

The author(s) declared no potential conficts of interest with respect to the research, authorship and/or publication of this article.

\section*{Data availability}
For reproducibility, all code and datasets are provided on GitHub \footnote{\href{https://github.com/ZhongLIFR/Logs2Graph}{https://github.com/ZhongLIFR/Logs2Graph}}.

\section*{Acknowledgements}
\textbf{Zhong Li} and \textbf{Matthijs van Leeuwen}: this publication is part of the project Digital Twin with project number P18-03 of the research programme TTW Perspective, which is (partly) financed by the Dutch Research Council (NWO). \textbf{Jiayang Shi}: This research is co-financed by the European Union H2020-MSCA-ITN-2020 under grant agreement no. 956172 (xCTing).

 %\bibliographystyle{elsarticle-num} 
 \bibliographystyle{abbrvnat}
 \bibliography{mybibliography}





% \end{thebibliography}
\end{document}